\documentclass[aps,prd,twocolumn,showpacs]{revtex4} 

\usepackage{graphicx}

\font\FermiSmallfont=cmssq8 scaled 1200

\def\LANLppthead#1{
\null 
\begin{center}\vskip -1.0truein{\hbox to 7.5truein {
\hfill
\vbox to 1in {\vfill \FermiSmallfont
              \hbox{#1}
              \vfill}
}}\vskip-0.0truein\end{center}}

\begin{document}

\title{Big Bang Nucleosynthesis with Independent Neutrino Distribution Functions}
\author{Christel J. Smith$^{1}$, George M.\ Fuller$^{1}$, Michael S. Smith$^{2}$}  
\affiliation{$^1$Department of Physics, University of California, San Diego, La Jolla, CA
92093-0319
,$^2$Physics Division, Oak Ridge National Laboratory, Oak Ridge, TN 37831-6354}   

\date{\today}

\begin{abstract}
We have performed new Big Bang Nucleosynthesis calculations which employ arbitrarily-specified, time-dependent neutrino and antineutrino distribution functions for each of up to four neutrino flavors. We self-consistently couple these distributions to the thermodynamics, the expansion rate and scale factor-time/temperature relationship, as well as to all relevant weak, electromagnetic, and strong nuclear reaction processes in the early universe.  With this approach, we can treat any scenario in which neutrino or antineutrino spectral distortion might arise.  These scenarios might include, for example,  decaying particles, active-sterile neutrino oscillations, and active-active neutrino oscillations in the presence of significant lepton numbers.  Our calculations allow lepton numbers and sterile neutrinos to be constrained with observationally-determined primordial helium and deuterium abundances.  We have modified a standard BBN code to perform these calculations and have made it available to the community. 
\end{abstract}
\maketitle

\section{Introduction}

There is a new paradigm in Big Bang Nucleosynthesis (BBN) studies which promises enhanced probes of the early universe and a window into new physics.  In the past, 
BBN predictions have been used to place constraints on the baryon number at three minutes after the Big Bang. This was done by comparing the observationally-inferred 
primordial light element abundances to abundances predicted by BBN calculations over a wide range of baryon-to-photon ratio values.  With the high precision results of the Wilkinson Microwave Anisotropy Probe (WMAP), however, the baryon-to-photon ratio, $\eta$, is now independently determined 
-- at 300,000 years after the Big Bang -- from observations of the cosmic microwave background (CMB) relative acoustic peak amplitudes \cite{WMAP, WMAP1, 3yrwmap}.  Currently, the WMAP Three Year Mean value for the baryon-to-photon ratio is $\eta = \left(6.11\pm .22\right) \times 10^{-10} $.  Future missions ($e.g.$, Planck\cite{bond}) promise considerably higher precision determinations of $\eta$.

Since the baryon-to-photon ratio is known independently, and to excellent precision albeit at much later times,  BBN calculations can now be used to probe or constrain new physics or heretofore poorly determined parameters.  For example, we can use BBN predictions to constrain not only the lepton numbers but also the physics behind these lepton numbers .  The existence of a nonzero electron lepton number follows from charge neutrality and the observed proton content of the universe.  The contributions of neutrinos and antineutrinos to the electron, muon, and tau $(e, \mu, \tau)$ lepton numbers are not known, since we do not directly observe these relic particles.  The neutrino contribution to the lepton number for a given flavor,  $\alpha={\rm e},\mu,\tau$, is defined analogously to the baryon-to-photon ratio, $\eta \equiv (n_b-n_{\bar b})/n_\gamma$, as
\begin{equation}
L_{\nu_\alpha} \equiv {{n_{\nu_\alpha}-n_{\bar\nu_\alpha}}\over{n_\gamma}},
\label{lepton}
\end{equation}
where $n_\gamma = (2\zeta(3)/\pi^2) T^3_\gamma$ is the proper photon number density at temperature $T_\gamma$, and $n_{\nu_\alpha}$ and $n_{\bar\nu_\alpha}$ are the neutrino and antineutrino number densities.  Observational bounds on the lepton numbers\cite{abfw, kfs, Kneller:2001cd, abb, wong, dolgov, Simha:2008mt, Cuoco:2003cu, Serpico:2005bc} remain large compared to the values of these that could significantly affect BBN when there is new leptonic sector physics ($e.g.$, sterile neutrinos)\cite{abfw}.

The neutrino lepton numbers influence BBN and the resulting primordial element abundances in a number of ways\cite{wfh}.  The energy density in the neutrino sector contributes to the total energy density of the universe which determines the expansion rate.  The expansion rate is crucial to the outcome of BBN because it determines the weak freeze-out temperature which in turn effectively sets the neutron-to-proton ratio and, therefore, the primordial abundances of $^4$He and the other light elements. 

Not only is the total number of neutrinos important to the outcome of BBN, but the neutrino distribution functions are key components of the phase space integrals in the weak reaction rates in BBN.  The weak reactions of greatest interest are those that inter-convert neutrons and protons:
\begin{equation}
 \nu_e+n\rightleftharpoons p+e^-,
  \label{nuen} 
 \end{equation}
 \begin{equation}
 \bar\nu_e+p\rightleftharpoons n+e^+, 
 \label{nuebarp}
 \end{equation}
 \begin{equation}
 n \rightleftharpoons p+e^-+\bar\nu_e. \label{ndecay}
\end{equation}
Since the rates for the weak reactions are strongly energy dependent, the energy distributions of the neutrinos and antineutrinos can figure prominently in both the forward and reverse rates in the processes in Eqs.~(\ref{nuen}), (\ref{nuebarp}), and (\ref{ndecay}).  In standard BBN scenarios the neutrino distribution functions are assumed to be thermally-shaped Fermi-Dirac distributions.  However, it is possible that non-thermal neutrino distribution functions arise after the neutrinos decouple from the background plasma around $T \approx 3\,{\rm MeV}$ and during times crucial to BBN.

There are many possible mechanisms that could alter the neutrino spectra. Altered neutrino energy spectra, in turn, could change the resulting primordial element abundances from what one would expect given a particular lepton number.  Neutrino energy spectrum-altering scenarios include, but are not limited to, active-active neutrino oscillations\cite{dolgov, abb, abfw, wong}, active-sterile neutrino oscillations\cite{abfw, kfs, sfka, cirelli, FV, fv97}, or particle decay into the neutrino sea\cite{pastor}.  Moreover, active-sterile neutrino flavor mixing and other mechanisms for creating sterile neutrino dark matter before neutrino decoupling are a focus of current research\cite{Dodelson:1993je, afp, Dolgov:2000ew, Shaposhnikov:2006xi, Kusenko:2006rh, Petraki:2007gq, Petraki:2008ef, Shi:1998fu, Chiu:1977ds, Boyanovsky:2007zz, Boyanovsky:2006it, Boyanovsky:2007ba, Abazajian:2002yz}, as is the constraint of these scenarios via x-ray observations and large-scale structure considerations\cite{Abazajian:2001vt, Abazajian:2006yn, Boyarsky:2006fg, Boyarsky:2005us, Yuksel:2007xh, Watson:2006qb, Viel:2005qj, Abazajian:2005xn, Seljak:2006qw}. Though these models may not directly affect BBN through the spectral distortion of $\nu_e$ and $\bar\nu_e$ energy distribution functions discussed here, they nevertheless may affect the overall values of lepton number, entropy, and energy density which are relevant to BBN.  In the end, the existence of sterile neutrino states changes the meaning and utility of lepton number\cite{Foot:1995qk, Shi:1996ic}. To use BBN predictions to probe or constrain any such scenario requires an approach that self-consistently includes neutrino and antineutrino energy spectra of arbitrary shape.

We have performed detailed  calculations of primordial nucleosynthesis in which we include neutrino and antineutrino spectral distortion.  Our results are surprising.  We find that even modest distortions of the neutrino and/or antineutrino spectral shapes from Fermi-Dirac black body forms can result in significant modification of the net neutron-proton interconversion rates and, hence, alteration of the light element abundances.

To study the effects of neutrino spectral distortion, we have modified the original Kawano/Wagoner BBN code described in Ref. \cite{skm} to calculate the primordial element abundances self-consistently with arbitrarily-specified non-thermal and/or time-dependent neutrino distribution functions.  This paper is  structured as follows: Section II describes the calculation of weak charge-changing reaction rates in the early universe and our prescription for employing non-thermal neutrino and antineutrino energy distribution functions; 
Section III discusses our new BBN code; 
Section IV will present example results for non-thermal neutrino distribution functions resulting from various physical scenarios; and Section V gives conclusions.

\section{BBN and the Weak Reaction Rates}

\begin{figure}
\includegraphics[width=2.5in,angle=270]{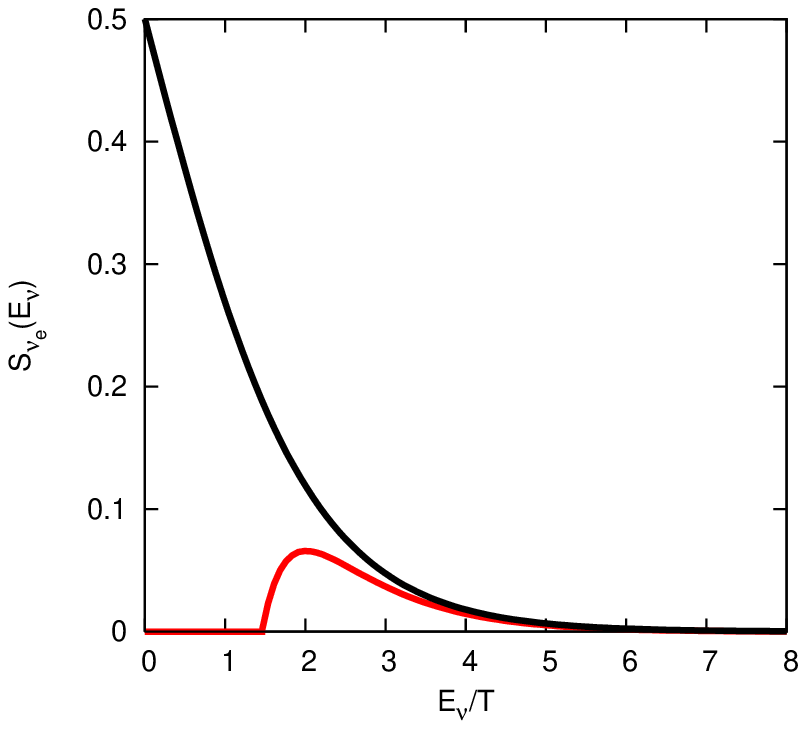}
\caption{Example neutrino occupation probabilities.  The upper dark (black) curve is the standard Fermi-Dirac thermally-distrubuted neutrino occupation probability and the lower light (red) curve is an example non-thermal neutrino occupation probability which can result from active-sterile neutrino transformation.}
\label{occprob}
\end{figure} 

At early times and high temperatures, $t\sim 1$ sec and $T\gtrsim 1$ MeV, the primordial element abundances are given by nuclear statistical equilibrium (NSE).  In NSE the rates for the processes which create a particular nucleus are equal to the rates that destroy it, so that the abundance for each element is given by the Saha equation.

As  the universe expands and cools, reaction rates slow down to the point where they will not be fast enough to maintain NSE and the 
neutron and proton abundances, and subsequently the abundances of $^4$He and the other light nuclei,  \textquotedblleft freeze-out".  For example, the $^4$He abundance falls below its equilibrium NSE track at $T\approx 0.6$ MeV, essentially as a consequence of the small NSE deuterium abundance.  BBN can be looked at crudely as a series of freeze-outs from NSE, but with considerable post-equilibrium nuclear processing.

Because the entropy per baryon is high, alpha particles form copiously during BBN.  Nearly all the neutrons in the universe at the epoch where $\alpha$'s form end up in alpha particles.

A key factor in the outcome of BBN is the value of the neutron-to-proton ratio.  Like the nuclear abundances in NSE, at high enough temperatures ($T >  3$~MeV) the {\it weak} neutron-proton inter-conversion rates are fast enough to maintain chemical equilibrium and the neutron-to-proton ratio can be determined from a Saha equation when the neutrinos have thermally-shaped distribution functions (as we will describe later).  

For general conditions the neutron-to-proton ratio is determined by the weak reaction processes shown in Eqs.\ (\ref{nuen}-\ref{ndecay}).  The rates for these weak reactions are given in Eqs.~(\ref{genep}-\ref{revrate}) below.  The forward rate for the reaction in Eq.\ (\ref{nuen}) is given by $\lambda_{\nu_en}$, Eq.~(\ref{nueonrate}), and the corresponding reverse rate is given by $\lambda_{e^-p}$, Eq.~(\ref{genep}).  Likewise, the forward and reverse rates for the process in Eq.\ (\ref{nuebarp}) are $\lambda_{\bar\nu_ep}$ and $\lambda_{e^+n}$ respectively.  Eq.\ (\ref{ndecayrate}) gives the rate for free neutron decay denoted by  $\lambda_{\rm n-decay}$, while the reverse three-body reaction rate is denoted by $\lambda_{pe^+\bar\nu_e}$ given in Eq.\ (\ref{revrate}). These rates are detailed below\cite{abfw, dicus, FFNI, FFNII, FFNIII, FFNIV}:

\begin{widetext}
\begin{equation}
 \lambda_{e^-p}  \approx {{ \ln{2}}\over{\langle ft\rangle { \left( m_ec^2 \right)}^5  }}
\int_{0}^\infty {{F \left[Z,E_\nu + Q_np\right] E_\nu^2 \left(E_\nu + Q_{np}\right)\left(\left(E_\nu + Q_{np}\right)^2 -m_ec^2\right)^{1/2} }\left[S_{e^-}\right] \left[ 1-{{S}}_{\nu_e} \right] dE_\nu  },\label{genep} 
\end{equation}

\begin{equation}
\lambda_{\bar\nu_e p} \approx {{ \ln{2}}\over{\langle ft\rangle {\left(m_ec^2 \right)}^5 }}
\int_{Q_{np} + m_ec^2}^\infty {{E_\nu ^2 \left( E_\nu - Q_{np} \right) \left(\left( E_\nu - Q_{np} \right)^2 -m_ec^2\right)^{1/2}}\left[S_{\bar\nu_e} \right] \left[1-S_{e^+}\right] dE_\nu}, \label{nuonprate} 
\end{equation}

\begin{equation}
\lambda_{e^+n} \approx {{ \ln{2}}\over{\langle ft\rangle {\left(m_ec^2 \right)}^5 }}
\int_{Q_{np} + m_ec^2}^\infty{{E_\nu^2 \left(E_\nu - Q_{np}\right)\left(\left(E_\nu - Q_{np}\right)^2 -m_ec^2\right)^{1/2}} \left[S_{e^+}\right] \left[1-S_{\bar\nu_e}\right] dE_\nu}, \label{eonnrate} 
\end{equation}

\begin{equation}
\lambda_{\nu_e n} \approx {{ \ln{2}}\over{\langle ft\rangle {\left(m_ec^2 \right)}^5 }}
\int_{0}^\infty{{F \left[Z,E_\nu + Q_np\right] E_\nu ^2 \left( E_\nu + Q_{np} \right) \left(\left( E_\nu + Q_{np} \right)^2 -m_ec^2\right)^{1/2}}\left[S_{\nu_e}\right] \left[1-S_{e^-}\right] dE_\nu}, \label{nueonrate}
\end{equation}

\begin {equation}
\lambda_{\rm n-decay} \approx {{ \ln{2}}\over{\langle ft\rangle {\left(m_ec^2 \right)}^5 }}
\int_{0}^{Q_{np}-m_ec^2} {F \left[Z,Q_np - E_\nu \right]E_\nu^2 \left( Q_{np}-E_\nu \right) \left( \left( Q_{np}-E_\nu \right)^2 -m_ec^2\right)^{1/2}} \left[1-S_{\bar\nu_e}\right] \left[1-S_{e^-} \right] dE_\nu , \label{ndecayrate}
\end{equation}

\begin{equation}
\lambda_{pe^+\bar\nu_e} \approx {{ \ln{2}}\over{\langle ft\rangle {\left(m_ec^2 \right)}^5 }}
\int_{0}^{Q_{np}-m_ec^2} {F \left[Z,Q_np - E_\nu \right] E_\nu^2 \left( Q_{np}-E_\nu \right) \left( \left( Q_{np}-E_\nu \right)^2 -m_ec^2\right)^{1/2}}  \left[S_{\bar\nu_e}\right] \left[S_{e^-} \right] dE_\nu ,\label{revrate}
\end{equation}
\end{widetext}
where $E_e$ and $E_\nu$ are the appropriate electron/positron and neutrino/antineutrino energies.  In these expressions the neutron-proton mass difference is $Q_{np} \approx 1.293$ MeV.  Here $\ln2/ \langle ft\rangle$ is proportional to the effective weak coupling applying to free nucleons with $\langle ft\rangle$ the effective $ft$-value defined in Ref.\cite{FFNI}.  The weak matrix element is $\ln2/ \langle ft\rangle \propto G_F^2(1+3g^2_A)$, where $G_F$ is the Fermi constant and $g_A$ is the ratio of axial to vector coupling for the free nucleons.  In the BBN calculation the value for $\ln2/ \langle ft\rangle$ is normalized by the free neutron decay  lifetime at zero-temperature.  Here $F\left[Z,E_e\right]$ is the relativistic coulomb correction factor (or Fermi factor)\cite{FFNI},
\begin{equation}
 F(\pm Z,w) \approx 2(1+s)(2pR)^{2(s-1)}e^{\pi\eta}\Bigg\vert{{\Gamma(s+i\eta)}\over{\Gamma(2s+1)}}\Bigg\vert.
 \label{coulomb}
\end{equation}  
In this expression the upper signs are for electron emission and capture, the lower signs are for positron emission and capture, $s=[1-(\alpha Z)^2]^{1/2}$, $Z$ is the appropriate nuclear charge (which is $Z=1$ for the proton), $\alpha$ is the fine structure constant, $\eta = \pm Zw/p$, and $R$ is the nuclear radius in electron Compton wavelengths.  $R\approx 2.908\times 10^{-3} A^{1/3} - 2.437A^{-1/3}$ where $A$ is the nuclear mass number and $\omega\equiv (p^2+m_e^2)^{1/2}$ with $m_e$ the electron rest mass.
This expression appears in the phase space integrand of the weak rates which require a Coulomb factor in either the initial or final state \cite{coulfac, dicus, L&T}.  

 $S_{e^-/+}$ and $S_{\nu_e/\bar\nu_e}$ are the phase space occupation probabilities for electrons/positrons and neutrinos/antineutrinos, respectively.  For example, the $\left[1-S_{\nu_e}\right]$ factor in $\lambda_{e^-p}$ is the Pauli phase space blocking factor for processes which create a neutrino.  In the limit that the neutrinos have {\it thermally-shaped} Fermi-Dirac distribution functions, these phase space occupation probabilities become two parameter functions:
\begin{equation}
S_{\nu_e} = {1\over { e^{E_{\nu_e}/{T_\nu} - \eta_{\nu_e}} +1}},
\label{nuocc}
\end{equation}
\begin{equation}
S_{\bar\nu_e} = {1 \over{ e^{E_{\nu_e}/{T_\nu} - \eta_{\bar\nu_e}} +1}}.
\label{nubarocc}
\end{equation}
The two parameters, $T_\nu$ and $\eta_{\nu_e}$, correspond to neutrino temperature and degeneracy parameter (the ratio of chemical potential to temperature), respectively.  For example, a thermally-shaped neutrino phase space occupation probability function is graphed in Fig.~\ref{occprob} as the upper black curve.

The total 
weak neutron destruction rate is $\lambda_n = \lambda_{\nu_e n} + \lambda_{e^+ n} + \lambda_{n-{\rm decay}}$ and the corresponding total weak proton destruction rate is $\lambda_p = \lambda_{\bar\nu_ep} + \lambda_{e^- p} + \lambda_{\bar\nu_e e^- p}$.  It is convenient to define 
\begin{equation}
\Lambda_{\rm tot}=\lambda_n +\lambda_p.
\label{total}
\end{equation}
With this definition, the rate of change of the net electron number per baryon, $Y_e$, with Friedmann-Lema$\hat{\rm i}$tre-Robertson-Walker (FLRW) time-like coordinate $t$ in the early universe will be
\begin{equation}
{{dY_e}\over{dt}}=\lambda_{n}-Y_e\, \Lambda_{\rm tot}.
\label{dyedt}
\end{equation} 

At early times where temperatures are high, the forward and reverse rates of these reactions are fast compared to the expansion rate of the universe. In this regime the neutron-to-proton ratio is just 
\begin{equation}
{{n}\over{p}}  = {{\lambda_{\bar\nu_ep}+\lambda_{e^-p}+\lambda_{pe\bar\nu_e}}\over{\lambda_{\nu_en}+\lambda_{e^+n}+\lambda_{n\ {\rm decay}}}}.
\label{ntopp}
\end{equation}
This can be approximated as
\begin{equation}
\label{ntoppp}
{{n}\over{p}}  \approx  {{\lambda_{\bar\nu_ep}+\lambda_{e^-p}}\over {\lambda_{\nu_en}+\lambda_{e^+n}}}
\end{equation}
because neutron decay and the reverse three-body reaction are negligible by comparison at high temperatures.  When the neutrino distribution functions have thermally-shaped Fermi-Dirac forms, the neutron-to-proton ratio is given by
\begin{equation}
\label{thernalnp}
{{n}\over{p}} \approx {{\left(\lambda_{e^-p}/\lambda_{e^+n}\right)+e^{-\eta_{\nu_e}+\eta_e-\xi}}\over{\left(\lambda_{e^-p}/\lambda_{e^+n}\right) e^{\eta_{\nu_e}-\eta_e+\xi}+1}},
\end{equation}
where $\eta_{\nu_e}=\mu_{\nu_e}/T$ is the electron neutrino degeneracy parameter, $\eta_e=\mu_e/T$ is the electron degeneracy parameter, and $\xi$ is the neutron-proton mass difference divided by temperature, $\xi= (m_n-m_p)/T$\cite{abfw}.  This equation is generally true whenever the lepton distribution functions have Fermi-Dirac forms and identical temperature parameters and whenever we can neglect neutron decay and its reverse process. Of course, at lower temperatures the neutrino and electron-photon plasma temperatures will differ and free neutron decay will be important.

\begin{figure}
\includegraphics[width=2.5in,angle=270]{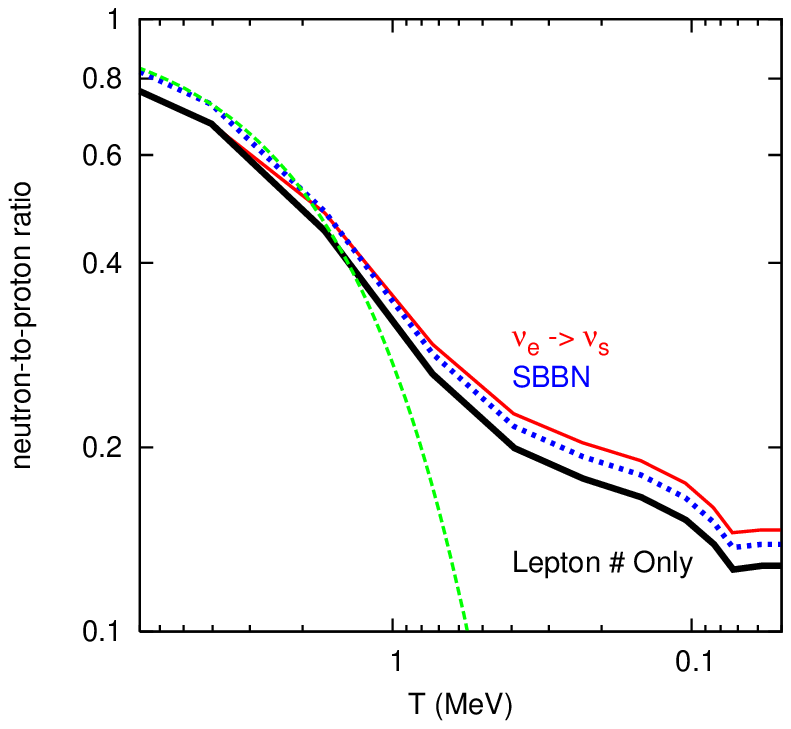}
\caption{The neutron to proton ratio, $n/p$, as a function of temperature for three nucleosynthesis scenarios.  The lower solid curve is for BBN with degenerate neutrinos and no neutrino transformation, where $L_{\nu_e} = L_{\nu_\tau} = L_{\nu_\mu} = .05$.  The upper solid curve is the $n/p$ ratio with the same lepton numbers as above but now including a particular active-sterile neutrino transformation scenario.  The dotted cure is the $n/p$ ratio for standard BBN (no lepton numbers or neutrino oscillation).  The dashed line is the $n/p$ equilibrium prediction for standard BBN (no lepton numbers or sterile neutrinos) with enforced weak chemical equilibrium.}
\label{ntopfig}
\end{figure}

If the weak reactions occur rapidly enough to maintain chemical equilibrium, then the Saha equation, $\mu_{\nu_e} + \mu_n = \mu_{e^-} + \mu_p$, can be used to predict the neutron-to-proton ratio.  Interestingly, both the Saha equation and the steady state rate equilibrium condition in Eq.\ (\ref{thernalnp}), with the full lepton capture rates of Eqs.~(\ref{genep}-\ref{revrate}), can be written as\cite{abfw}
\begin{equation}
\label{chemnp}
{{n}\over{p}} \approx e^{\left({\mu_e-\mu_{\nu_e}-\delta m_{np}}\right)/{T}}.
\end{equation}
This equilibrium neutron-to-proton ratio is shown in Fig.~\ref{ntopfig} as the dashed (green) line for zero electron and neutron chemical potentials, $\mu_e=\mu_{\nu_e} =0$.

As the universe cools, the weak reaction rates become slow compared to the expansion of the universe and the neutron-to-proton ratio falls out of equilibrium.  This is called \textquotedblleft weak freeze-out" and occurs over a range of  
temperatures.  Fig.~\ref{ntopfig} shows the actual neutron-to-proton ratio evolving as a function of temperature for the standard BBN scenario (thermal neutrino distribution functions and zero chemical potentials $\mu_e=\mu_{\nu_e} =0$).  At high temperatures, the actual neutron-to-proton ratio follows the equilibrium value and then around 1 MeV, the weak freeze-out commences.  This happens because the weak rates have a stronger dependence on temperature than does the expansion rate of the universe.  The lepton capture/decay rates given in Eqs.~(\ref{genep}-\ref{revrate}) scale very roughly as $T^5$ (see Ref.\cite{FFNIV} for the detailed temperature dependence), while the expansion rate of the universe is $\propto T^2$.  As a result, the neutron-proton weak interconversion rates eventually will fall below the expansion rate.

Although the weak rates become relatively slow, they still have a significant effect on the neutron-to-proton ratio, even for temperatures well below $T = 0.8$ MeV.  In fact, free neutron decay continues to lower the $n/p$ ratio until there are virtually no more free neutrons or until the neutrons are sequestered in alpha particles, where they are effectively shielded from the weak interaction.  This is illustrated in Fig.~\ref{ntopfig} where the dotted (blue) line continues to decrease until $T \approx .08$ MeV (when the neutrons have been captured during rapid alpha particle formation). It is important to correctly calculate the weak reactions in order to appropriately track the $n/p$ ratio. This ratio sets the scale, in varying degrees, for all the primordial element abundances\cite{skm, wfh}.
 
\section{New BBN Code}

A nucleosynthesis code was written by Robert V. Wagoner in 1969\cite{wag73, wag69} to track and time evolve the nuclear abundances and the neutron-to-proton ratio in an expanding cooling universe. It was later updated and revised by Lawrence Kawano in 1988\cite{kawano1}. 

This code time-evolves three main quantities, the electron fraction, $Y_e$, the baryon-to-photon ratio, $\eta$, and the temperature, along with the primordial element abundances.  It follows 48 nuclides using a reaction network composed of 168 nuclear reactions, whose rates have primarily been based on, and in some cases extrapolated from, laboratory cross sections.  The main numerical technique is a 2nd order Runga-Kutta routine.  

The code also tracks the neutron-to-proton ratio by calculating the weak reaction rates using the standard thermally-shaped Fermi-Dirac neutrino distribution functions, setting $S_{\nu_e}$ and $S_{\bar\nu_e}$ as given in Eq.~(\ref{nuocc}) and Eq.~(\ref{nubarocc}).

In their approach, electron energy is used as the integration variable, instead of neutrino energy as given in  Eqs.~(\ref{genep}-\ref{revrate}) above.  To save computational time, they calculate only the sum of each of the  forward $n\rightarrow p$ rates and the reverse $p\rightarrow n$ rates:
\begin{equation}
\lambda_{n} = \lambda_{\nu_e+n\rightarrow p+e^-} + \lambda_{n+e^+\rightarrow p+\bar\nu_e} + \lambda_{n\rightarrow p+e^-+\bar\nu_e}
\label{n-rates}
\end {equation}
\begin{equation}
\lambda_p = \lambda_{p+e^-\rightarrow \nu_e+n} + \lambda_{\bar\nu_e+p\rightarrow n+e^+} + \lambda_{p+e^-+\bar\nu_e\rightarrow n}.
\label{p-rates}
\end{equation}
With an algebraic trick, this simplifies the calculation by condensing the six phase space integrals (for each weak reaction rate) into two integrals:
\begin{widetext}

\begin{eqnarray}
\label{ntotwag}
\lambda_n &  \approx & {{\ln{2}}\over {\langle ft \rangle {\left(m_ec^2 \right)}^5 }} 
\\
& \times & \int_{m_ec^2}^{\infty}  E_e\left( E_e^2 -\left(m_ec^2\right)^2\right)^{1/2} \left[ {{\left(E_e+Q_{np}\right)^2}\over{\left(e^{E_e/T} +1\right) \left( e^{-\left(E_e+Q_{np}\right)/T_\nu -\eta_{\nu_e}}+1\right)}} +{{\left(E_e-Q_{np}\right)^2}\over{\left(e^{-E_e/T} +1\right) \left(e^{\left(E_e-Q_{np}\right)/T_{\nu} -\eta_{\nu_e}} +1\right)}}\right]dE_e
\nonumber
\end{eqnarray}

\begin{eqnarray}
\label{ptotwag}
\lambda_p & \approx & {{\ln{2}}\over {\langle ft \rangle {\left(m_ec^2 \right)}^5 }}
\\
 & \times &  \int_{m_ec^2}^{\infty} E_e\left( E_e^2 -\left(m_ec^2\right)^2\right)^{1/2} \left[ {{\left(E_e+Q_{np}\right)^2} \over {\left ( e^{E_e/T} +1\right) \left(e^{\left (E_e+ Q_{np} \right)/{T_\nu} + \eta_{\nu_e}} +1 \right)}} + {{\left(Q_{np} - E_e\right)^2} \over{ \left( e^{E_e/T} +1\right) \left( e^{\left (Q_{np} - E_e \right) /T_\nu +\eta_{\nu_e} }+1\right)}} \right] dE_e.
\nonumber
\end{eqnarray}
\end{widetext}
This algebraic trick requires the approximation of thermally-shaped Fermi-Dirac neutrino and antineutrino distribution functions. This summed rate cannot properly treat the Coulomb correction, $F[Z, E_e]$, which should be included in the phase space integral of reaction rates which have an electron and proton in either the final or initial state.  
\begin{figure}
\includegraphics[width=2.5in,angle=270]{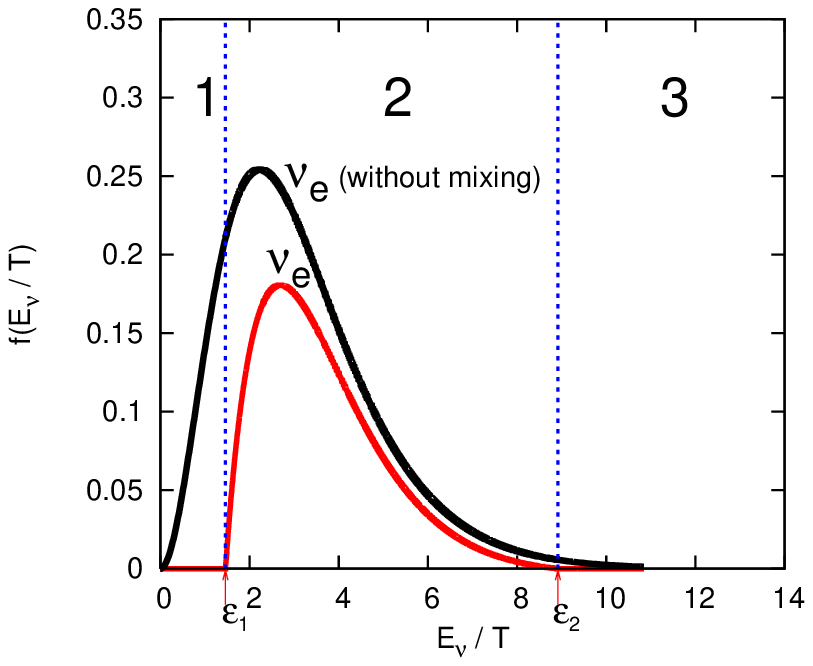}
\caption{Two example electron neutrino distribution functions, where the upper black line is the standard thermal spectrum and the lower red line is a spectrum resulting from a particular scenario for active-sterile neutrino mixing.  The vertical dashed lines show where a weak rate calculation employing the lower distribution function would be broken up to be integrated piece-wise in our new version of the code.}
\label{nudist}
\end{figure}

We have modified the Kawano/Wagoner BBN code so that it can accommodate and integrate any arbitrary neutrino and/or antineutrino distribution function with any specified time dependence.  The 
majority of our changes lie in the weak reaction rate calculation. 

We first separated the summed neutron destruction and production rates, $\lambda_n$ and $\lambda_p$.  This 
enabled us to use non-thermal distribution functions and to change the neutrino and antineutrino distribution functions independently.  Then, we removed 
a series approximation for $\lambda_n$ and $\lambda_p$ which is applied when the lepton numbers are zero.  This approximation results in an erroneous $\approx 0.5\%$ increase in the neutron-to-proton ratio\cite{kawano,kawano1}. Furthermore, we added the capability to separate a weak rate calculation into an arbitrary number of neutrino energy bins.  This is useful for calculating a reaction rate where the neutrino energy spectrum is comprised of different functions over different energy ranges. 

For example, in Fig.~\ref{nudist}, we have shown two electron neutrino distribution functions.  The upper curve is just the standard thermally-shaped Fermi-Dirac distribution function, 
\begin{equation}
\label{nudistributionfunction}
f_{\nu_\alpha}(E_\nu) = {{1}\over{T_{\nu_\alpha}^3 F_2\left(\eta_{\nu_\alpha}\right)}}{{{E_{\nu}}^2}\over{e^{E_\nu/T_{\nu_\alpha}-\eta_{\nu_\alpha}}+1}}, 
\end{equation}
which is consistent with the occupation probability  derived from Eq.\ (\ref{nuocc}). The lower curve is a distribution function resulting from a particular active-sterile neutrino oscillation scheme described in Refs.~\cite{kfs, sfka}.  In this scheme, electron neutrinos have been completely converted into steriles at low and high energies (1 and 3), leaving only active neutrinos in the center (2) energy band.   To calculate a rate using this non-thermal distribution function, we break up the rate into three parts.   The first part integrates from zero to $\epsilon_1$ using the neutrino distribution function $f(E_\nu /T) = 0$.  The second part integrates from $\epsilon_1$ to $\epsilon_2$ using the modified function shown in 2.  The third part integrates from $\epsilon_2$ to $\infty$ and again use $f(E_\nu /T) = 0$.  Finally, the total rate is calculated by summing all three pieces.

\begin{figure}
\includegraphics[width=3.5in,angle=0]{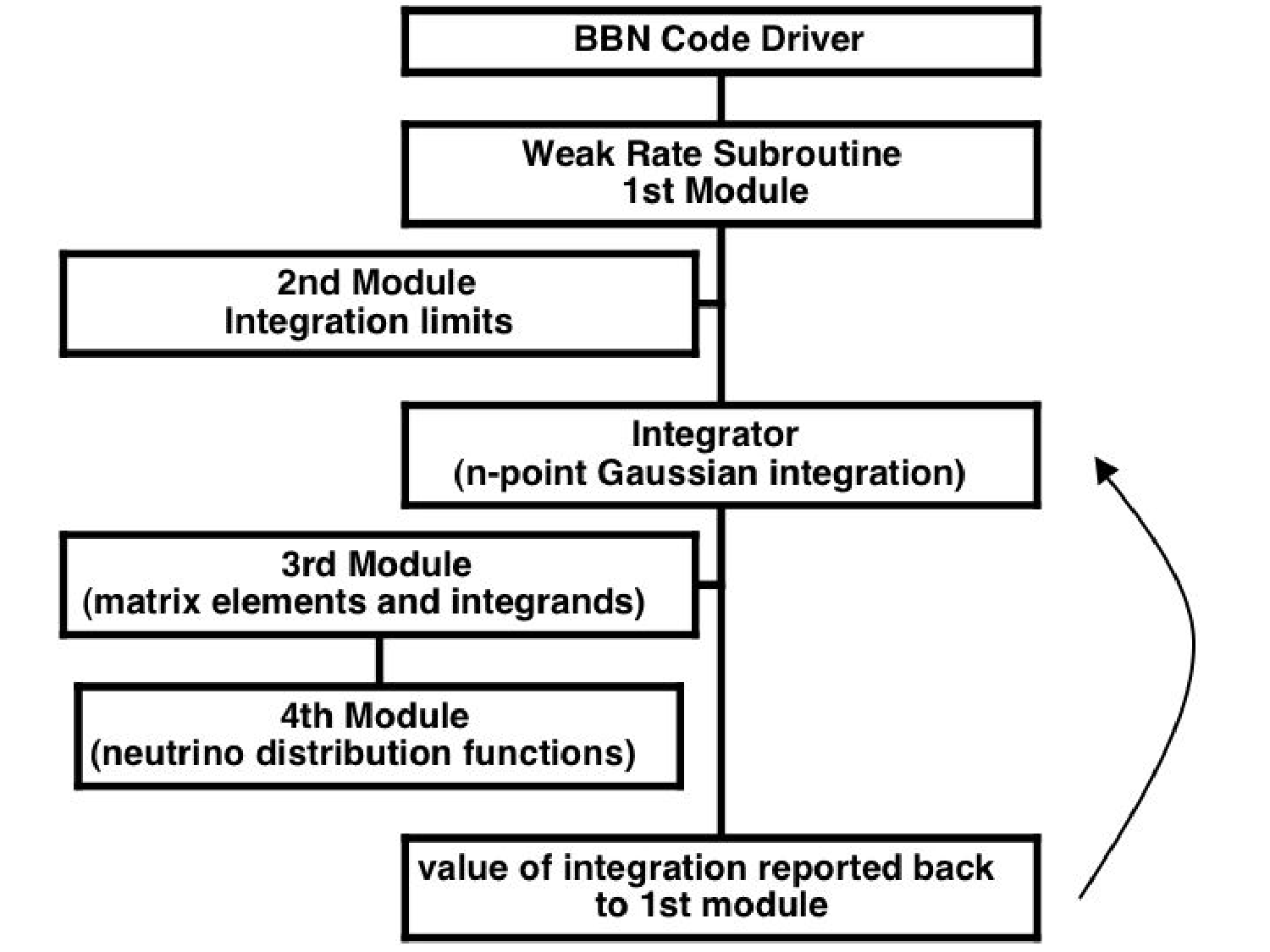}
\caption{Flow chart for our modified BBN calculation.}
\label{flowchart}
\end{figure}

To perform these non-thermal piece-wise calculations in the BBN code, we completely replaced the original weak rate calculation with a series of four modules.  These modules allow the user to define the distribution functions, break up the integration into specifiable pieces and define the energy ranges for each piece, and set any desired time/temperature dependence of the distribution functions. A flow chart of the weak rate calculations is shown in Fig.~\ref{flowchart}.  At each time step, the BBN code calls the weak rate calculation subroutine, Module 1 in Fig.~\ref{flowchart}, to time-evolve the neutron to proton ratio and, subsequently, all the nuclear abundances.  

Module 1 acts as the central line of communication in that it calls the other modules and reports back the value of the weak rates at every time step in the BBN code.   In this module, the user can first define how many pieces to split the rate integration into for reactions involving either neutrinos or antineutrinos or both.  For example, if the user wanted to use the lower non-thermal neutrino distribution function in Fig.~\ref{nudist} and a thermal antineutrino distribution function, the user can specify that the rate integrations involving neutrinos should be integrated in three parts and that rates involving antineutrinos should be integrated with one energy bin.  

Next, Module 1 calls Module 2 to retrieve the integration limits for each piece, $i.e.$, where the user wants each energy bin to begin and end.   In Module 2, the user can define these integration limits and couple them to any time dependences desired.  Module 1 makes an array with these limits so they can be accessed later in the integration.  This procedure can be extended to an arbitrary number of energy bins for any neutrino type.

The first module calculates all six weak reaction rates by utilizing two main loops. These loop over the number of energy bins.  One loop calculates the two reaction rates that include neutrinos and the other loop calculates the four remaining weak reaction rates that include antineutrinos.  The number of iterations for each loop is determined by the number of energy bins.  Each loop iteration integrates the weak reaction rates over the range of energy and neutrino distribution function specified for that energy bin.  At the end of the iteration, each rate is summed.

For every loop cycle, the first module calls the integrator which inputs the function to be integrated and the limits of the energy bins (from Module 2). The matrix elements and integrands for the six weak reaction rates, as shown in Eqs.~(\ref{genep}-\ref{revrate}), are retrieved from Module 3. Here, the electron occupation probability is set as $S_e=1/(e^{E_e/T} +1)$ and the neutrino and antineutrino occupation probabilities are called from Module 4. 

The sole purpose of Module 4 is to house the neutrino and antineutrino occupation probabilities.  This makes it easy for a user to modify the neutrino distribution functions 
-- by inputting analytic functions for $S_{\nu_e}$ and $S_{\bar\nu_e}$ --  without having to modify any other portion of the weak rate calculation.  The user can also define different functions or populations for each integration energy bin.  After each energy bin is integrated, the total rate is summed and the values for the six weak reaction rates are returned to the main BBN code driver.

Our modified Kawano/Wagoner BBN code -- which can now accommodate and integrate any arbitrary neutrino and/or antineutrino distribution function with any specified time dependence -- will be available to the community at bigbangonline.org\cite{bigbangonline}. 

\section{Example Code Results}

\begin{figure}
\includegraphics[width=2.5in,angle=270]{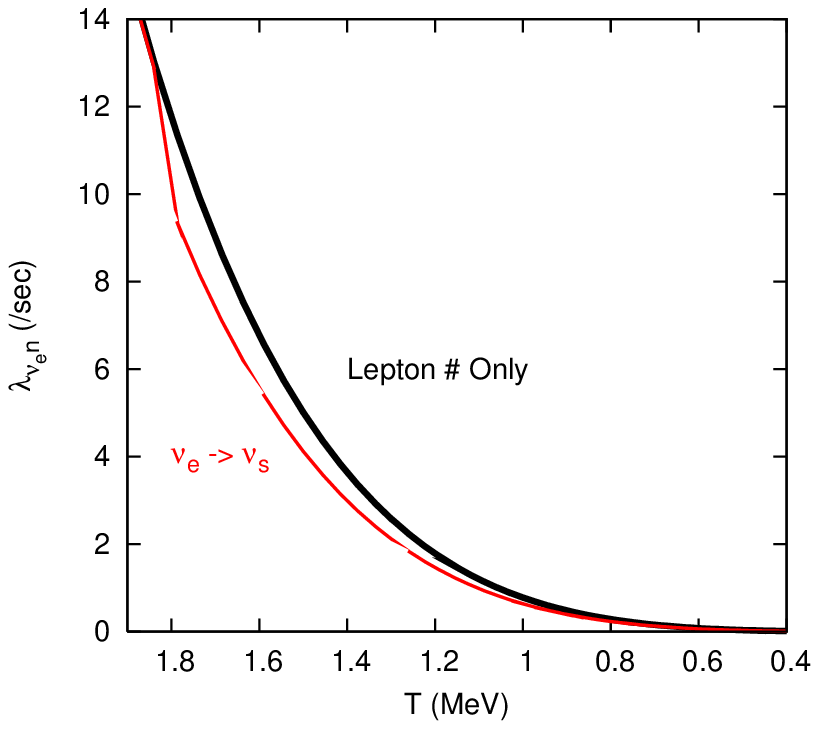}
\caption{The rate of electron neutrino capture on a neutron as a function of temperature.  The upper curve is $\lambda_{\nu_e {\rm n}}$ in the lepton number only case for lepton numbers of  $L_{\nu_e} = L_{\nu_\tau} = L_{\nu_\mu} = .05$.  The lower curve is the rate when there is active-sterile neutrino transformation along with the same lepton numbers as above.}
\label{nuonnfig}
\end{figure}

We have utilized this code to study nucleosynthesis abundance yields in the presence of a light-mass sterile neutrino over a range of lepton numbers\cite{kfs, sfka}.  The lower red line in Fig.~\ref{occprob} shows a final non-thermal neutrino occupation probability function that can result from active-sterile neutrino transformation.  In this particular scenario, we started with normal thermal electron neutrino and antineutrino distribution functions and an assumed initial lepton number. The lepton numbers that we have taken are within the range which is allowed by conventional BBN (primordial $^4{\rm He}$) considerations. But, of course, the point is that a sterile neutrino which mixes with an active neutrino can result in non-thermal neutrino and/or antineutrino energy spectra which produce BBN abundance yields which can be quite different than in the standard scenario. This, in turn, could provide new, more appropriate constraints on lepton numbers or on active-sterile neutrino mass and mixing parameter space or on both. 

The presence of a significant net lepton number can delay significant sterile neutrino production until after the weak decoupling temperature.  With a positive net lepton number, a Mikheyev-Smirnov-Wolfenstein (MSW) resonance occurs first for low neutrino energies. This resonance subsequently sweeps to higher neutrino energies as the universe expands and cools.  At first, this resonance sweep process occurs adiabatically, efficiently  converting all active neutrinos into sterile neutrinos. This continues until the rate of active-sterile conversion becomes too fast to maintain adiabaticity.  At this point, production becomes inefficient. However, at high enough resonance energies transformations can occur adiabatically again.  

Accurately following such a scenario requires all the modifications in our new code. Without being able to include a dynamically changing neutrino distribution function, for example, we could not calculate correctly the neutron-to-proton inter-conversion rates. In fact, in the example scenario presented here, not only are there non-thermal neutrino distribution functions to handle, but these change on time scales which are important to BBN.  In Fig.~\ref{nuonnfig}, we show the rate for electron neutrino capture on a neutron, the forward process in Eq.~\ref{nuen}, as a function of temperature.  The top curve is the rate when there is no active-sterile neutrino oscillation.  The lower curve shows the decreased rate when there is active-sterile mixing and the final neutrino distribution function is that of Fig.~\ref{occprob}. By reducing the number of electron neutrinos available for capture on neutrons, the capture rate is decreased.  Additionally, the altered neutrino distribution function also results in a modestly increased reverse rate (electron capture on protons). The depleted electron neutrino distribution function in this scenario has the effect of increasing the electron capture rate because of the smaller neutrino phase space blocking factor.  

The final integrated effect in this scenario can be gauged by the changes in the light element abundances.  For example, with adopted lepton numbers of $L_{\nu_e}=L_{\nu_\mu}=L_{\nu_\tau} =  0.05$, which corresponds to a electron, mu, and tau neutrino degeneracy parameters of, $\eta_{\nu_e}=\eta_{\nu_\mu}=\eta_{\nu_\tau} \approx 0.073$ ({\it i.e.,} near the conventional BBN upper limits on these quantities), we see a $4.9\%$ increase of $^4$He over the standard (no neutrino mixing and no lepton numbers) BBN value and a $12.7\%$ increase over the $^4$He calculation with only lepton numbers included but no active-sterile neutrino oscillation effects.  With this example scenario we find an increase in D/H (deuterium abundance relative to hydrogen) of $2.8\%$ over the standard BBN calculation and an increase of $6.9\%$ from the lepton number only calculation. 

The increase in helium for these adopted parameters is likely unacceptable, exceeding observational bounds\cite{OS, Olive, IT}. Likewise, if the observationally-determined value of D/H can be increased in precision sufficiently (to better than $\pm 5 \%$ \cite{sfka}), it may be possible that D/H could compete with helium as an avenue for constraint of new neutrino physics. Ultimately, allowing for dynamically-altered neutrino and antineutrino distribution functions could add a new dimension to the way in which BBN and light element abundances might constrain new physics in the weak sector.

We have also used our new code to apply a relativistic version of the Coulomb correction into the appropriate weak rate integrands\cite{coulfac}.  This has never been done before in the Wagoner/Kawano BBN code.

\section{conclusion}

We have developed an approach to Big Bang Nucleosynthesis (BBN) calculations where we can treat arbitrarily-specified energy distributions for all neutrino types, including $\nu_e$ and $\bar\nu_e$.  We can also allow these distribution functions to be altered dynamically and follow all nuclear and weak reactions self-consistently with these alterations. This new approach can extend the usefulness of BBN predictions for exploring and constraining new physics in the neutrino and weak interaction sectors.  

Examples of such new physics include active-sterile neutrino mixing and particle decays that have neutrinos in the final state.  We have given an explicit example of the former scenario. In this example we have demonstrated how active-sterile neutrino oscillation physics can alter neutrino or antineutrino distribution functions on short time scales, alter the neutron-proton inter-conversions rates, and so modify BBN abundance yields over those of the standard scenario. 

Our calculations hold out the promise that light element abundances could place the best constraints on primordial lepton numbers and active-sterile neutrino mixing parameters when the sterile neutrino mass is in the $\sim 1\,{\rm eV}$ range. Present laboratory experiments, like mini-BooNE, are sensitive to neutrino flavor mixing in the active-sterile channel at the $\sim 1,{\rm eV}$ mass scale only when the appropriate effective $2\times 2$ vacuum mixing angle satisfies $\sin^22\theta \gg {10}^{-4}$. By contrast, in the presence of a net lepton number, BBN abundance yields might be significantly altered for  active-sterile neutrino mixing parameters for $\sin^22\theta > {10}^{-8}$. The greater reach in vacuum mixing angle afforded by BBN considerations stems from: (1) the long (gravitational) expansion time scale of the early universe which dictates the MSW resonance sweep rate and sets the minimum mixing angle required for adiabatic and efficient conversion of the active neutrinos into sterile species; and (2) the significant sensitivity of the neutron-proton weak inter-conversion rates to alterations of the neutrino or antineutrino energy distribution functions. Our new calculations allow us to follow simultaneously and self-consistently both of these effects along with all relevant weak, electromagnetic, and strong nuclear reaction rates.

This new approach is incorporated into an update of the Kawano/Wagoner BBN code -- which can now accommodate and integrate any arbitrary neutrino and/or antineutrino distribution function with any specified time dependence.  
We will soon make this code available to the community at bigbangonline.org.

\begin{acknowledgments}
We would like to acknowledge discussions with Chad Kishimoto and Kevork Abazajian.  ORNL is managed by UT-Battelle, LLC, for the U.S. DOE under contract DE-AC05-00OR22725.  The work of G.M.F and C.J.S. was supported in part by a NSF grant  and a UC/LANL CARE grant at UCSD.
\end{acknowledgments}

\bibliography{mybiblio}

\begin{thebibliography}{59}
\expandafter\ifx\csname natexlab\endcsname\relax\def\natexlab#1{#1}\fi
\expandafter\ifx\csname bibnamefont\endcsname\relax
  \def\bibnamefont#1{#1}\fi
\expandafter\ifx\csname bibfnamefont\endcsname\relax
  \def\bibfnamefont#1{#1}\fi
\expandafter\ifx\csname citenamefont\endcsname\relax
  \def\citenamefont#1{#1}\fi
\expandafter\ifx\csname url\endcsname\relax
  \def\url#1{\texttt{#1}}\fi
\expandafter\ifx\csname urlprefix\endcsname\relax\def\urlprefix{URL }\fi
\providecommand{\bibinfo}[2]{#2}
\providecommand{\eprint}[2][]{\url{#2}}

\bibitem[{\citenamefont{Tegmark et~al.}(2004)}]{WMAP}
\bibinfo{author}{\bibfnamefont{M.}~\bibnamefont{Tegmark}} \bibnamefont{et~al.}
  (\bibinfo{collaboration}{SDSS Collaboration}), \bibinfo{journal}{Phys.\ Rev.\
  D} \textbf{\bibinfo{volume}{69}}, \bibinfo{pages}{103501}
  (\bibinfo{year}{2004}).

\bibitem[{\citenamefont{Spergel et~al.}(2003)}]{WMAP1}
\bibinfo{author}{\bibfnamefont{D.~N.} \bibnamefont{Spergel}}
  \bibnamefont{et~al.}, \bibinfo{journal}{Astrophys.\ J.\ Suppl.}
  \textbf{\bibinfo{volume}{148}}, \bibinfo{pages}{175} (\bibinfo{year}{2003}).

\bibitem[{\citenamefont{Spergel et~al.}(2006)}]{3yrwmap}
\bibinfo{author}{\bibfnamefont{D.~N.} \bibnamefont{Spergel}}
  \bibnamefont{et~al.} (\bibinfo{year}{2006}), \eprint{astro-ph/0603449}.

\bibitem[{\citenamefont{Bond et~al.}(2004)\citenamefont{Bond, Contaldi, Lewis,
  and Pogosyan}}]{bond}
\bibinfo{author}{\bibfnamefont{J.~R.} \bibnamefont{Bond}},
  \bibinfo{author}{\bibfnamefont{C.}~\bibnamefont{Contaldi}},
  \bibinfo{author}{\bibfnamefont{A.}~\bibnamefont{Lewis}}, \bibnamefont{and}
  \bibinfo{author}{\bibfnamefont{D.}~\bibnamefont{Pogosyan}},
  \bibinfo{journal}{Int.\ J.\ Theor.\ Phys.} \textbf{\bibinfo{volume}{43}},
  \bibinfo{pages}{599} (\bibinfo{year}{2004}).

\bibitem[{\citenamefont{Abazajian et~al.}(2005)\citenamefont{Abazajian, Bell,
  Fuller, and Wong}}]{abfw}
\bibinfo{author}{\bibfnamefont{K.}~\bibnamefont{Abazajian}},
  \bibinfo{author}{\bibfnamefont{N.~F.} \bibnamefont{Bell}},
  \bibinfo{author}{\bibfnamefont{G.~M.} \bibnamefont{Fuller}},
  \bibnamefont{and} \bibinfo{author}{\bibfnamefont{Y.~Y.~Y.}
  \bibnamefont{Wong}}, \bibinfo{journal}{Phys.\ Rev.\ D}
  \textbf{\bibinfo{volume}{72}}, \bibinfo{pages}{063004}
  (\bibinfo{year}{2005}).

\bibitem[{\citenamefont{Kishimoto et~al.}(2006)\citenamefont{Kishimoto, Fuller,
  and Smith}}]{kfs}
\bibinfo{author}{\bibfnamefont{C.~T.} \bibnamefont{Kishimoto}},
  \bibinfo{author}{\bibfnamefont{G.~M.} \bibnamefont{Fuller}},
  \bibnamefont{and} \bibinfo{author}{\bibfnamefont{C.~J.} \bibnamefont{Smith}},
  \bibinfo{journal}{Phys. Rev. Lett.} \textbf{\bibinfo{volume}{97}},
  \bibinfo{pages}{141301} (\bibinfo{year}{2006}), \eprint{astro-ph/0607403}.

\bibitem[{\citenamefont{Kneller et~al.}(2001)\citenamefont{Kneller, Scherrer,
  Steigman, and Walker}}]{Kneller:2001cd}
\bibinfo{author}{\bibfnamefont{J.~P.} \bibnamefont{Kneller}},
  \bibinfo{author}{\bibfnamefont{R.~J.} \bibnamefont{Scherrer}},
  \bibinfo{author}{\bibfnamefont{G.}~\bibnamefont{Steigman}}, \bibnamefont{and}
  \bibinfo{author}{\bibfnamefont{T.~P.} \bibnamefont{Walker}},
  \bibinfo{journal}{Phys. Rev.} \textbf{\bibinfo{volume}{D64}},
  \bibinfo{pages}{123506} (\bibinfo{year}{2001}), \eprint{astro-ph/0101386}.

\bibitem[{\citenamefont{Abazajian et~al.}(2002)\citenamefont{Abazajian, Beacom,
  and Bell}}]{abb}
\bibinfo{author}{\bibfnamefont{K.~N.} \bibnamefont{Abazajian}},
  \bibinfo{author}{\bibfnamefont{J.~F.} \bibnamefont{Beacom}},
  \bibnamefont{and} \bibinfo{author}{\bibfnamefont{N.~F.} \bibnamefont{Bell}},
  \bibinfo{journal}{Phys.\ Rev.\ D} \textbf{\bibinfo{volume}{66}},
  \bibinfo{pages}{013008} (\bibinfo{year}{2002}).

\bibitem[{\citenamefont{Wong}(2002)}]{wong}
\bibinfo{author}{\bibfnamefont{Y.~Y.~Y.} \bibnamefont{Wong}},
  \bibinfo{journal}{Phys.\ Rev.\ D} \textbf{\bibinfo{volume}{66}},
  \bibinfo{pages}{025015} (\bibinfo{year}{2002}).

\bibitem[{\citenamefont{Dolgov et~al.}(2002)\citenamefont{Dolgov, Hansen,
  Pastor, Petcov, Raffelt, and Semikoz}}]{dolgov}
\bibinfo{author}{\bibfnamefont{A.~D.} \bibnamefont{Dolgov}},
  \bibinfo{author}{\bibfnamefont{S.~H.} \bibnamefont{Hansen}},
  \bibinfo{author}{\bibfnamefont{S.}~\bibnamefont{Pastor}},
  \bibinfo{author}{\bibfnamefont{S.~T.} \bibnamefont{Petcov}},
  \bibinfo{author}{\bibfnamefont{G.~G.} \bibnamefont{Raffelt}},
  \bibnamefont{and} \bibinfo{author}{\bibfnamefont{D.~V.}
  \bibnamefont{Semikoz}}, \bibinfo{journal}{Nucl.\ Phys.\ B}
  \textbf{\bibinfo{volume}{632}}, \bibinfo{pages}{363} (\bibinfo{year}{2002}).

\bibitem[{\citenamefont{Simha and Steigman}(2008)}]{Simha:2008mt}
\bibinfo{author}{\bibfnamefont{V.}~\bibnamefont{Simha}} \bibnamefont{and}
  \bibinfo{author}{\bibfnamefont{G.}~\bibnamefont{Steigman}},
  \bibinfo{journal}{JCAP} \textbf{\bibinfo{volume}{0808}}, \bibinfo{pages}{011}
  (\bibinfo{year}{2008}), \eprint{hep-ph/0806.0179}.

\bibitem[{\citenamefont{Cuoco et~al.}(2004)}]{Cuoco:2003cu}
\bibinfo{author}{\bibfnamefont{A.}~\bibnamefont{Cuoco}} \bibnamefont{et~al.},
  \bibinfo{journal}{Int. J. Mod. Phys.} \textbf{\bibinfo{volume}{A19}},
  \bibinfo{pages}{4431} (\bibinfo{year}{2004}), \eprint{astro-ph/0307213}.

\bibitem[{\citenamefont{Serpico and Raffelt}(2005)}]{Serpico:2005bc}
\bibinfo{author}{\bibfnamefont{P.~D.} \bibnamefont{Serpico}} \bibnamefont{and}
  \bibinfo{author}{\bibfnamefont{G.~G.} \bibnamefont{Raffelt}},
  \bibinfo{journal}{Phys. Rev.} \textbf{\bibinfo{volume}{D71}},
  \bibinfo{pages}{127301} (\bibinfo{year}{2005}), \eprint{astro-ph/0506162}.

\bibitem[{\citenamefont{Wagoner et~al.}(1967)\citenamefont{Wagoner, Fowler, and
  Hoyle}}]{wfh}
\bibinfo{author}{\bibfnamefont{R.~V.} \bibnamefont{Wagoner}},
  \bibinfo{author}{\bibfnamefont{W.~A.} \bibnamefont{Fowler}},
  \bibnamefont{and} \bibinfo{author}{\bibfnamefont{F.}~\bibnamefont{Hoyle}},
  \bibinfo{journal}{Astrophys.\ J.} \textbf{\bibinfo{volume}{148}},
  \bibinfo{pages}{3} (\bibinfo{year}{1967}).

\bibitem[{\citenamefont{Smith et~al.}(2006)\citenamefont{Smith, Fuller,
  Kishimoto, and Abazajian}}]{sfka}
\bibinfo{author}{\bibfnamefont{C.~J.} \bibnamefont{Smith}},
  \bibinfo{author}{\bibfnamefont{G.~M.} \bibnamefont{Fuller}},
  \bibinfo{author}{\bibfnamefont{C.~T.} \bibnamefont{Kishimoto}},
  \bibnamefont{and} \bibinfo{author}{\bibfnamefont{K.~N.}
  \bibnamefont{Abazajian}}, \bibinfo{journal}{Phys. Rev.}
  \textbf{\bibinfo{volume}{D74}}, \bibinfo{pages}{085008}
  (\bibinfo{year}{2006}), \eprint{astro-ph/0608377}.

\bibitem[{\citenamefont{Chu and Cirelli}(2006)}]{cirelli}
\bibinfo{author}{\bibfnamefont{Y.-Z.} \bibnamefont{Chu}} \bibnamefont{and}
  \bibinfo{author}{\bibfnamefont{M.}~\bibnamefont{Cirelli}},
  \bibinfo{journal}{Phys. Rev.} \textbf{\bibinfo{volume}{D74}},
  \bibinfo{pages}{085015} (\bibinfo{year}{2006}), \eprint{astro-ph/0608206}.

\bibitem[{\citenamefont{Foot and Volkas}(1995)}]{FV}
\bibinfo{author}{\bibfnamefont{R.}~\bibnamefont{Foot}} \bibnamefont{and}
  \bibinfo{author}{\bibfnamefont{R.~R.} \bibnamefont{Volkas}},
  \bibinfo{journal}{Phys.\ Rev.\ Lett.} \textbf{\bibinfo{volume}{75}},
  \bibinfo{pages}{4350} (\bibinfo{year}{1995}).

\bibitem[{\citenamefont{Foot and Volkas}(1997)}]{fv97}
\bibinfo{author}{\bibfnamefont{R.}~\bibnamefont{Foot}} \bibnamefont{and}
  \bibinfo{author}{\bibfnamefont{R.~R.} \bibnamefont{Volkas}},
  \bibinfo{journal}{Phys.\ Rev.\ D} \textbf{\bibinfo{volume}{55}},
  \bibinfo{pages}{5147} (\bibinfo{year}{1997}).

\bibitem[{\citenamefont{Cuoco et~al.}(2005)\citenamefont{Cuoco, Lesgourgues,
  Mangano, and Pastor}}]{pastor}
\bibinfo{author}{\bibfnamefont{A.}~\bibnamefont{Cuoco}},
  \bibinfo{author}{\bibfnamefont{J.}~\bibnamefont{Lesgourgues}},
  \bibinfo{author}{\bibfnamefont{G.}~\bibnamefont{Mangano}}, \bibnamefont{and}
  \bibinfo{author}{\bibfnamefont{S.}~\bibnamefont{Pastor}},
  \bibinfo{journal}{Phys. Rev.} \textbf{\bibinfo{volume}{D71}},
  \bibinfo{pages}{123501} (\bibinfo{year}{2005}), \eprint{astro-ph/0502465}.

\bibitem[{\citenamefont{Dodelson and Widrow}(1994)}]{Dodelson:1993je}
\bibinfo{author}{\bibfnamefont{S.}~\bibnamefont{Dodelson}} \bibnamefont{and}
  \bibinfo{author}{\bibfnamefont{L.~M.} \bibnamefont{Widrow}},
  \bibinfo{journal}{Phys. Rev. Lett.} \textbf{\bibinfo{volume}{72}},
  \bibinfo{pages}{17} (\bibinfo{year}{1994}), \eprint{hep-ph/9303287}.

\bibitem[{\citenamefont{Abazajian
  et~al.}(2001{\natexlab{a}})\citenamefont{Abazajian, Fuller, and Patel}}]{afp}
\bibinfo{author}{\bibfnamefont{K.}~\bibnamefont{Abazajian}},
  \bibinfo{author}{\bibfnamefont{G.~M.} \bibnamefont{Fuller}},
  \bibnamefont{and} \bibinfo{author}{\bibfnamefont{M.}~\bibnamefont{Patel}},
  \bibinfo{journal}{Phys. Rev.} \textbf{\bibinfo{volume}{D64}},
  \bibinfo{pages}{023501} (\bibinfo{year}{2001}{\natexlab{a}}),
  \eprint{astro-ph/0101524}.

\bibitem[{\citenamefont{Dolgov and Hansen}(2002)}]{Dolgov:2000ew}
\bibinfo{author}{\bibfnamefont{A.~D.} \bibnamefont{Dolgov}} \bibnamefont{and}
  \bibinfo{author}{\bibfnamefont{S.~H.} \bibnamefont{Hansen}},
  \bibinfo{journal}{Astropart. Phys.} \textbf{\bibinfo{volume}{16}},
  \bibinfo{pages}{339} (\bibinfo{year}{2002}), \eprint{hep-ph/0009083}.

\bibitem[{\citenamefont{Shaposhnikov and Tkachev}(2006)}]{Shaposhnikov:2006xi}
\bibinfo{author}{\bibfnamefont{M.}~\bibnamefont{Shaposhnikov}}
  \bibnamefont{and} \bibinfo{author}{\bibfnamefont{I.}~\bibnamefont{Tkachev}},
  \bibinfo{journal}{Phys. Lett.} \textbf{\bibinfo{volume}{B639}},
  \bibinfo{pages}{414} (\bibinfo{year}{2006}), \eprint{hep-ph/0604236}.

\bibitem[{\citenamefont{Kusenko}(2006)}]{Kusenko:2006rh}
\bibinfo{author}{\bibfnamefont{A.}~\bibnamefont{Kusenko}},
  \bibinfo{journal}{Phys. Rev. Lett.} \textbf{\bibinfo{volume}{97}},
  \bibinfo{pages}{241301} (\bibinfo{year}{2006}), \eprint{hep-ph/0609081}.

\bibitem[{\citenamefont{Petraki and Kusenko}(2008)}]{Petraki:2007gq}
\bibinfo{author}{\bibfnamefont{K.}~\bibnamefont{Petraki}} \bibnamefont{and}
  \bibinfo{author}{\bibfnamefont{A.}~\bibnamefont{Kusenko}},
  \bibinfo{journal}{Phys. Rev.} \textbf{\bibinfo{volume}{D77}},
  \bibinfo{pages}{065014} (\bibinfo{year}{2008}), \eprint{astro-ph/0711.4646}.

\bibitem[{\citenamefont{Petraki}(2008)}]{Petraki:2008ef}
\bibinfo{author}{\bibfnamefont{K.}~\bibnamefont{Petraki}},
  \bibinfo{journal}{Phys. Rev.} \textbf{\bibinfo{volume}{D77}},
  \bibinfo{pages}{105004} (\bibinfo{year}{2008}), \eprint{hep-ph/0801.3470}.

\bibitem[{\citenamefont{Shi and Fuller}(1999)}]{Shi:1998fu}
\bibinfo{author}{\bibfnamefont{X.-D.} \bibnamefont{Shi}} \bibnamefont{and}
  \bibinfo{author}{\bibfnamefont{G.~M.} \bibnamefont{Fuller}},
  \bibinfo{journal}{Phys. Rev. Lett.} \textbf{\bibinfo{volume}{83}},
  \bibinfo{pages}{3120} (\bibinfo{year}{1999}), \eprint{astro-ph/9904041}.

\bibitem[{\citenamefont{Chiu et~al.}(1977)\citenamefont{Chiu, Sudarshan, and
  Misra}}]{Chiu:1977ds}
\bibinfo{author}{\bibfnamefont{C.~B.} \bibnamefont{Chiu}},
  \bibinfo{author}{\bibfnamefont{E.~C.~G.} \bibnamefont{Sudarshan}},
  \bibnamefont{and} \bibinfo{author}{\bibfnamefont{B.}~\bibnamefont{Misra}},
  \bibinfo{journal}{Phys. Rev.} \textbf{\bibinfo{volume}{D16}},
  \bibinfo{pages}{520} (\bibinfo{year}{1977}).

\bibitem[{\citenamefont{Boyanovsky and
  Ho}(2007{\natexlab{a}})}]{Boyanovsky:2007zz}
\bibinfo{author}{\bibfnamefont{D.}~\bibnamefont{Boyanovsky}} \bibnamefont{and}
  \bibinfo{author}{\bibfnamefont{C.~M.} \bibnamefont{Ho}},
  \bibinfo{journal}{Phys. Rev.} \textbf{\bibinfo{volume}{D76}},
  \bibinfo{pages}{085011} (\bibinfo{year}{2007}{\natexlab{a}}),
  \eprint{hep-ph/0705.0703}.

\bibitem[{\citenamefont{Boyanovsky and
  Ho}(2007{\natexlab{b}})}]{Boyanovsky:2006it}
\bibinfo{author}{\bibfnamefont{D.}~\bibnamefont{Boyanovsky}} \bibnamefont{and}
  \bibinfo{author}{\bibfnamefont{C.~M.} \bibnamefont{Ho}},
  \bibinfo{journal}{JHEP} \textbf{\bibinfo{volume}{07}}, \bibinfo{pages}{030}
  (\bibinfo{year}{2007}{\natexlab{b}}), \eprint{hep-ph/0612092}.

\bibitem[{\citenamefont{Boyanovsky}(2008)}]{Boyanovsky:2007ba}
\bibinfo{author}{\bibfnamefont{D.}~\bibnamefont{Boyanovsky}},
  \bibinfo{journal}{Phys. Rev.} \textbf{\bibinfo{volume}{D77}},
  \bibinfo{pages}{023528} (\bibinfo{year}{2008}), \eprint{astro-ph/0711.0470}.

\bibitem[{\citenamefont{Abazajian and Fuller}(2002)}]{Abazajian:2002yz}
\bibinfo{author}{\bibfnamefont{K.~N.} \bibnamefont{Abazajian}}
  \bibnamefont{and} \bibinfo{author}{\bibfnamefont{G.~M.}
  \bibnamefont{Fuller}}, \bibinfo{journal}{Phys. Rev.}
  \textbf{\bibinfo{volume}{D66}}, \bibinfo{pages}{023526}
  (\bibinfo{year}{2002}), \eprint{astro-ph/0204293}.

\bibitem[{\citenamefont{Abazajian
  et~al.}(2001{\natexlab{b}})\citenamefont{Abazajian, Fuller, and
  Tucker}}]{Abazajian:2001vt}
\bibinfo{author}{\bibfnamefont{K.}~\bibnamefont{Abazajian}},
  \bibinfo{author}{\bibfnamefont{G.~M.} \bibnamefont{Fuller}},
  \bibnamefont{and} \bibinfo{author}{\bibfnamefont{W.~H.}
  \bibnamefont{Tucker}}, \bibinfo{journal}{Astrophys. J.}
  \textbf{\bibinfo{volume}{562}}, \bibinfo{pages}{593}
  (\bibinfo{year}{2001}{\natexlab{b}}), \eprint{astro-ph/0106002}.

\bibitem[{\citenamefont{Abazajian and Koushiappas}(2006)}]{Abazajian:2006yn}
\bibinfo{author}{\bibfnamefont{K.}~\bibnamefont{Abazajian}} \bibnamefont{and}
  \bibinfo{author}{\bibfnamefont{S.~M.} \bibnamefont{Koushiappas}},
  \bibinfo{journal}{Phys. Rev.} \textbf{\bibinfo{volume}{D74}},
  \bibinfo{pages}{023527} (\bibinfo{year}{2006}), \eprint{astro-ph/0605271}.

\bibitem[{\citenamefont{Boyarsky
  et~al.}(2006{\natexlab{a}})\citenamefont{Boyarsky, Neronov, Ruchayskiy,
  Shaposhnikov, and Tkachev}}]{Boyarsky:2006fg}
\bibinfo{author}{\bibfnamefont{A.}~\bibnamefont{Boyarsky}},
  \bibinfo{author}{\bibfnamefont{A.}~\bibnamefont{Neronov}},
  \bibinfo{author}{\bibfnamefont{O.}~\bibnamefont{Ruchayskiy}},
  \bibinfo{author}{\bibfnamefont{M.}~\bibnamefont{Shaposhnikov}},
  \bibnamefont{and} \bibinfo{author}{\bibfnamefont{I.}~\bibnamefont{Tkachev}},
  \bibinfo{journal}{Phys. Rev. Lett.} \textbf{\bibinfo{volume}{97}},
  \bibinfo{pages}{261302} (\bibinfo{year}{2006}{\natexlab{a}}),
  \eprint{astro-ph/0603660}.

\bibitem[{\citenamefont{Boyarsky
  et~al.}(2006{\natexlab{b}})\citenamefont{Boyarsky, Neronov, Ruchayskiy, and
  Shaposhnikov}}]{Boyarsky:2005us}
\bibinfo{author}{\bibfnamefont{A.}~\bibnamefont{Boyarsky}},
  \bibinfo{author}{\bibfnamefont{A.}~\bibnamefont{Neronov}},
  \bibinfo{author}{\bibfnamefont{O.}~\bibnamefont{Ruchayskiy}},
  \bibnamefont{and}
  \bibinfo{author}{\bibfnamefont{M.}~\bibnamefont{Shaposhnikov}},
  \bibinfo{journal}{Mon. Not. Roy. Astron. Soc.}
  \textbf{\bibinfo{volume}{370}}, \bibinfo{pages}{213}
  (\bibinfo{year}{2006}{\natexlab{b}}), \eprint{astro-ph/0512509}.

\bibitem[{\citenamefont{Yuksel et~al.}(2008)\citenamefont{Yuksel, Beacom, and
  Watson}}]{Yuksel:2007xh}
\bibinfo{author}{\bibfnamefont{H.}~\bibnamefont{Yuksel}},
  \bibinfo{author}{\bibfnamefont{J.~F.} \bibnamefont{Beacom}},
  \bibnamefont{and} \bibinfo{author}{\bibfnamefont{C.~R.}
  \bibnamefont{Watson}}, \bibinfo{journal}{Phys. Rev. Lett.}
  \textbf{\bibinfo{volume}{101}}, \bibinfo{pages}{121301}
  (\bibinfo{year}{2008}), \eprint{astro-ph/0706.4084}.

\bibitem[{\citenamefont{Watson et~al.}(2006)\citenamefont{Watson, Beacom,
  Yuksel, and Walker}}]{Watson:2006qb}
\bibinfo{author}{\bibfnamefont{C.~R.} \bibnamefont{Watson}},
  \bibinfo{author}{\bibfnamefont{J.~F.} \bibnamefont{Beacom}},
  \bibinfo{author}{\bibfnamefont{H.}~\bibnamefont{Yuksel}}, \bibnamefont{and}
  \bibinfo{author}{\bibfnamefont{T.~P.} \bibnamefont{Walker}},
  \bibinfo{journal}{Phys. Rev.} \textbf{\bibinfo{volume}{D74}},
  \bibinfo{pages}{033009} (\bibinfo{year}{2006}), \eprint{astro-ph/0605424}.

\bibitem[{\citenamefont{Viel et~al.}(2005)\citenamefont{Viel, Lesgourgues,
  Haehnelt, Matarrese, and Riotto}}]{Viel:2005qj}
\bibinfo{author}{\bibfnamefont{M.}~\bibnamefont{Viel}},
  \bibinfo{author}{\bibfnamefont{J.}~\bibnamefont{Lesgourgues}},
  \bibinfo{author}{\bibfnamefont{M.~G.} \bibnamefont{Haehnelt}},
  \bibinfo{author}{\bibfnamefont{S.}~\bibnamefont{Matarrese}},
  \bibnamefont{and} \bibinfo{author}{\bibfnamefont{A.}~\bibnamefont{Riotto}},
  \bibinfo{journal}{Phys. Rev.} \textbf{\bibinfo{volume}{D71}},
  \bibinfo{pages}{063534} (\bibinfo{year}{2005}), \eprint{astro-ph/0501562}.

\bibitem[{\citenamefont{Abazajian}(2006)}]{Abazajian:2005xn}
\bibinfo{author}{\bibfnamefont{K.}~\bibnamefont{Abazajian}},
  \bibinfo{journal}{Phys. Rev.} \textbf{\bibinfo{volume}{D73}},
  \bibinfo{pages}{063513} (\bibinfo{year}{2006}), \eprint{astro-ph/0512631}.

\bibitem[{\citenamefont{Seljak et~al.}(2006)\citenamefont{Seljak, Makarov,
  McDonald, and Trac}}]{Seljak:2006qw}
\bibinfo{author}{\bibfnamefont{U.}~\bibnamefont{Seljak}},
  \bibinfo{author}{\bibfnamefont{A.}~\bibnamefont{Makarov}},
  \bibinfo{author}{\bibfnamefont{P.}~\bibnamefont{McDonald}}, \bibnamefont{and}
  \bibinfo{author}{\bibfnamefont{H.}~\bibnamefont{Trac}},
  \bibinfo{journal}{Phys. Rev. Lett.} \textbf{\bibinfo{volume}{97}},
  \bibinfo{pages}{191303} (\bibinfo{year}{2006}), \eprint{astro-ph/0602430}.

\bibitem[{\citenamefont{Foot et~al.}(1996)\citenamefont{Foot, Thomson, and
  Volkas}}]{Foot:1995qk}
\bibinfo{author}{\bibfnamefont{R.}~\bibnamefont{Foot}},
  \bibinfo{author}{\bibfnamefont{M.~J.} \bibnamefont{Thomson}},
  \bibnamefont{and} \bibinfo{author}{\bibfnamefont{R.~R.}
  \bibnamefont{Volkas}}, \bibinfo{journal}{Phys. Rev.}
  \textbf{\bibinfo{volume}{D53}}, \bibinfo{pages}{5349} (\bibinfo{year}{1996}),
  \eprint{hep-ph/9509327}.

\bibitem[{\citenamefont{Shi}(1996)}]{Shi:1996ic}
\bibinfo{author}{\bibfnamefont{X.-D.} \bibnamefont{Shi}},
  \bibinfo{journal}{Phys. Rev.} \textbf{\bibinfo{volume}{D54}},
  \bibinfo{pages}{2753} (\bibinfo{year}{1996}), \eprint{astro-ph/9602135}.

\bibitem[{\citenamefont{Smith et~al.}(1993)\citenamefont{Smith, Kawano, and
  Malaney}}]{skm}
\bibinfo{author}{\bibfnamefont{M.~S.} \bibnamefont{Smith}},
  \bibinfo{author}{\bibfnamefont{L.~H.} \bibnamefont{Kawano}},
  \bibnamefont{and} \bibinfo{author}{\bibfnamefont{R.~A.}
  \bibnamefont{Malaney}}, \bibinfo{journal}{Astrophys.\ J.\ Suppl.}
  \textbf{\bibinfo{volume}{85}}, \bibinfo{pages}{219} (\bibinfo{year}{1993}).

\bibitem[{\citenamefont{Dicus et~al.}(1982)}]{dicus}
\bibinfo{author}{\bibfnamefont{D.~A.} \bibnamefont{Dicus}}
  \bibnamefont{et~al.}, \bibinfo{journal}{Phys. Rev.}
  \textbf{\bibinfo{volume}{D26}}, \bibinfo{pages}{2694} (\bibinfo{year}{1982}).

\bibitem[{\citenamefont{Fuller et~al.}(1980)\citenamefont{Fuller, Fowler, and
  Newman}}]{FFNI}
\bibinfo{author}{\bibfnamefont{G.~M.} \bibnamefont{Fuller}},
  \bibinfo{author}{\bibfnamefont{W.~A.} \bibnamefont{Fowler}},
  \bibnamefont{and} \bibinfo{author}{\bibfnamefont{M.~J.}
  \bibnamefont{Newman}}, \bibinfo{journal}{Astrophys.\ J.\ Suppl.}
  \textbf{\bibinfo{volume}{42}}, \bibinfo{pages}{447} (\bibinfo{year}{1980}).

\bibitem[{\citenamefont{Fuller et~al.}(1982{\natexlab{a}})\citenamefont{Fuller,
  Fowler, and Newman}}]{FFNII}
\bibinfo{author}{\bibfnamefont{G.~M.} \bibnamefont{Fuller}},
  \bibinfo{author}{\bibfnamefont{W.~A.} \bibnamefont{Fowler}},
  \bibnamefont{and} \bibinfo{author}{\bibfnamefont{M.~J.}
  \bibnamefont{Newman}}, \bibinfo{journal}{Astrophys. J.}
  \textbf{\bibinfo{volume}{252}}, \bibinfo{pages}{715}
  (\bibinfo{year}{1982}{\natexlab{a}}).

\bibitem[{\citenamefont{Fuller et~al.}(1982{\natexlab{b}})\citenamefont{Fuller,
  Fowler, and Newman}}]{FFNIII}
\bibinfo{author}{\bibfnamefont{G.~M.} \bibnamefont{Fuller}},
  \bibinfo{author}{\bibfnamefont{W.~A.} \bibnamefont{Fowler}},
  \bibnamefont{and} \bibinfo{author}{\bibfnamefont{M.~J.}
  \bibnamefont{Newman}}, \bibinfo{journal}{Astrophys. J. Suppl.}
  \textbf{\bibinfo{volume}{48}}, \bibinfo{pages}{279}
  (\bibinfo{year}{1982}{\natexlab{b}}).

\bibitem[{\citenamefont{{Fuller} et~al.}(1985)\citenamefont{{Fuller}, {Fowler},
  and {Newman}}}]{FFNIV}
\bibinfo{author}{\bibfnamefont{G.~M.} \bibnamefont{{Fuller}}},
  \bibinfo{author}{\bibfnamefont{W.~A.} \bibnamefont{{Fowler}}},
  \bibnamefont{and} \bibinfo{author}{\bibfnamefont{M.~J.}
  \bibnamefont{{Newman}}}, \bibinfo{journal}{\apj}
  \textbf{\bibinfo{volume}{293}}, \bibinfo{pages}{1} (\bibinfo{year}{1985}).

\bibitem[{\citenamefont{Smith and Fuller}(2008)}]{coulfac}
\bibinfo{author}{\bibfnamefont{C.~J.} \bibnamefont{Smith}} \bibnamefont{and}
  \bibinfo{author}{\bibfnamefont{G.~M.} \bibnamefont{Fuller}},
  \bibinfo{journal}{In preparation}  (\bibinfo{year}{2008}).

\bibitem[{\citenamefont{Lopez and Turner}(1999)}]{L&T}
\bibinfo{author}{\bibfnamefont{R.~E.} \bibnamefont{Lopez}} \bibnamefont{and}
  \bibinfo{author}{\bibfnamefont{M.~S.} \bibnamefont{Turner}},
  \bibinfo{journal}{Phys. Rev.} \textbf{\bibinfo{volume}{D59}},
  \bibinfo{pages}{103502} (\bibinfo{year}{1999}), \eprint{astro-ph/9807279}.

\bibitem[{\citenamefont{Wagoner}(1973)}]{wag73}
\bibinfo{author}{\bibfnamefont{R.~V.} \bibnamefont{Wagoner}},
  \bibinfo{journal}{Astrophys. J.} \textbf{\bibinfo{volume}{179}},
  \bibinfo{pages}{343} (\bibinfo{year}{1973}).

\bibitem[{\citenamefont{Wagoner}(1969)}]{wag69}
\bibinfo{author}{\bibfnamefont{R.~V.} \bibnamefont{Wagoner}},
  \bibinfo{journal}{Ann. Rev. Astron. Astrophys.} \textbf{\bibinfo{volume}{7}},
  \bibinfo{pages}{553} (\bibinfo{year}{1969}).

\bibitem[{\citenamefont{Kawano}(1988)}]{kawano1}
\bibinfo{author}{\bibfnamefont{L.}~\bibnamefont{Kawano}}
  (\bibinfo{year}{1988}), \eprint{FERMILAB-PUB-88/34-A}.

\bibitem[{\citenamefont{Kawano}(1992)}]{kawano}
\bibinfo{author}{\bibfnamefont{L.}~\bibnamefont{Kawano}},
  \bibinfo{journal}{NASA STI/Recon Technical Report N}
  \textbf{\bibinfo{volume}{92}}, \bibinfo{pages}{25163} (\bibinfo{year}{1992}).

\bibitem[{\citenamefont{Smith}()}]{bigbangonline}
\bibinfo{author}{\bibfnamefont{M.~S.} \bibnamefont{Smith}},
  \emph{\bibinfo{title}{Big bang online}},
  \bibinfo{howpublished}{\url{http://bigbangonline.org}}.

\bibitem[{\citenamefont{Olive and Skillman}(2004)}]{OS}
\bibinfo{author}{\bibfnamefont{K.~A.} \bibnamefont{Olive}} \bibnamefont{and}
  \bibinfo{author}{\bibfnamefont{E.~D.} \bibnamefont{Skillman}},
  \bibinfo{journal}{Astrophys.\ J.} \textbf{\bibinfo{volume}{617}},
  \bibinfo{pages}{29} (\bibinfo{year}{2004}).

\bibitem[{\citenamefont{Olive et~al.}(1997)\citenamefont{Olive, Steigman, and
  Skillman}}]{Olive}
\bibinfo{author}{\bibfnamefont{K.~A.} \bibnamefont{Olive}},
  \bibinfo{author}{\bibfnamefont{G.}~\bibnamefont{Steigman}}, \bibnamefont{and}
  \bibinfo{author}{\bibfnamefont{E.~D.} \bibnamefont{Skillman}},
  \bibinfo{journal}{Astrophys.\ J.} \textbf{\bibinfo{volume}{483}},
  \bibinfo{pages}{788} (\bibinfo{year}{1997}).

\bibitem[{\citenamefont{Izotov and Thuan}(2004)}]{IT}
\bibinfo{author}{\bibfnamefont{Y.~I.} \bibnamefont{Izotov}} \bibnamefont{and}
  \bibinfo{author}{\bibfnamefont{T.~X.} \bibnamefont{Thuan}},
  \bibinfo{journal}{Astrophys.\ J.} \textbf{\bibinfo{volume}{602}},
  \bibinfo{pages}{200} (\bibinfo{year}{2004}).

\end{thebibliography}

\end{document}